\documentclass[conference]{IEEEtran}
\IEEEoverridecommandlockouts
\usepackage{cite}
\usepackage{amsmath,amssymb,amsfonts}
\usepackage{algorithmic}
\usepackage{graphicx}
\usepackage{textcomp}
\usepackage{xcolor}
\usepackage{multirow}
\def\BibTeX{{\rm B\kern-.05em{\sc i\kern-.025em b}\kern-.08em
    T\kern-.1667em\lower.7ex\hbox{E}\kern-.125emX}}
\usepackage{hyperref}
    
\begin{document}

\title{A Decentralized Resource Management System Proposal For Disasters: NGO-RMSD (STK-AKYS)}

\author{\IEEEauthorblockN{1\textsuperscript{st} Arzu Özkan}
\IEEEauthorblockA{\textit{Department of English Language \& Literature} \\
\textit{Muğla Sıtkı Koçman University}\\
Muğla, Turkey \\
arzuozknn8@gmail.com}
\and
\IEEEauthorblockN{2\textsuperscript{nd} Umutcan Korkmaz}
\IEEEauthorblockA{\textit{Department of Computer Engineering} \\
\textit{Muğla Sıtkı Koçman University}\\
Muğla, Turkey \\
kmmzurtkcn@gmail.com}
\and
\IEEEauthorblockN{3\textsuperscript{rd} Cemal Dak}
\IEEEauthorblockA{\textit{Department of Computer Engineering} \\
\textit{Muğla Sıtkı Koçman University}\\
Muğla, Turkey \\
cemal.dak@gmail.com}
\and
\IEEEauthorblockN{4\textsuperscript{th} Enis Karaarslan}
\IEEEauthorblockA{\textit{Department of Computer Engineering} \\
\textit{Muğla Sıtkı Koçman University}\\
Muğla, Turkey \\
enis.karaarslan@mu.edu.tr}
}

\maketitle

\begin{abstract}
Disaster and emergency management are under the responsibility of many organizations and there are serious coordination problems in post-disaster crisis management. This paper proposes a decentralized non-governmental organization resource management system for disasters (NGO-RMSD). This system is based on blockchain technology and it will enable the non-governmental organizations (NGO) and public institutions to manage and coordinate the resources in a trusted environment in the case of disasters. A proof of concept implementation is developed by using the Quorum blockchain framework which is more energy-efficient than crypto currency-based blockchain solutions. Smart contracts are developed for the autonomous working of the system. These smart contacts are used for the verification of the needs of the one who is in need, delivering resources to the right people, and identifying the urgent needs. The system aims to reach more disaster victims in a more timely manner. NGO-RMSD is designed according to the needs of the NGOs in the field. The application is shared with the free software license and further development with the community is aimed.

\end{abstract}

\begin{IEEEkeywords}
Disaster, Non-Governmental Organizations, Crisis management, Organization, Source management, Coordination, Blockchain
\end{IEEEkeywords}

\section{Introduction}
The disaster management processes of the local governments are not sufficient even in developed countries such as Japan and America [1]. Disaster and emergency management are under the responsibility of many organizations, there are critical coordination problems in post-disaster crisis management. 

Non-governmental organizations and volunteers have significant roles and contributions in the case of disasters. Today, NGOs intensely work on the following about the disasters: 
\begin{itemize}
\item Support the preparation of communities for the disasters
\item Work on the reduction of pre-disaster risks and post-disaster losses, 
\item Work on short-term intervention, and long-term improvement efforts. 
\end{itemize}

Without the help of voluntary organizations, it does not seem possible for local governments and the state to meet the needs of disaster victims. There is a need for Joint disaster management that is possible by assigning the roles of voluntary organizations at all stages of the disasters and establishing coordination between parties [2]. 

The duty of NGOs is to support local governments and the public, as needed especially in times of crisis within the framework of their chosen mission [3]. As an example; Ahbap NGO took an active role in the post-disaster process after the 6.8 magnitude earthquake in Elazig Sivrice/TURKEY, which happened in January 2020. One of our team members, who was an active Ahbap volunteer in this aid team, observed the following:
\begin{itemize}
\item Non-governmental organizations working in the disaster area act independently of each other
\item There is a deficiency and problem to reach the real needy people 
\item The resources can not be used efficiently. 
\end{itemize}

When NGOs are analyzed, it is seen that there are a variety of NGOs which have different missions and visions. Each has various resources and strengths. The most significant deficiency that we have observed in the post-disaster management is; the coordination problem between NGOs and the government. This results in the efficient usage of the capabilities.  This situation causes deficiencies in resource management and slows down the post-disaster recovery efforts.
    
In the next section, the fundamentals will be given. Then the related works will be given in Section 3. In Section 4, details of the system proposal are given.  The proof of concept implementation of NGO-RMSD system is given in Section 5. Conclusion and future works will be given in the last section.

\section{Fundamentals}
\IEEEpubidadjcol

\subsection{Blockchain}

Blockchain is a distributed ledger technology that ensures trust between the users without an intermediary. This is accomplished by keeping track of records linked to each other[4]. The records are called transactions and transactions are bundled into blocks at the nodes (the computers that are in charge of the blockchain system). These records are stored in immutable record storage called the ledger.  
Since the blockchain system is distributed, there is a need for a way that the nodes can agree upon the transactions, blocks, etc. Consensus protocols are used in blockchain networks to provide that functionality to the distributed system. The consensus protocol in the blockchain is chosen according to the type of the blockchain.

Different types of blockchain networks exist. Their types mainly depend on who can join as a node and participate in the network and also the visibility of the blockchain records. Public/private types are according to the visibility of the records. In public blockchain networks; all the transactions in the network are visible by anyone on the internet. Private networks only allow authorized users to reach the records. Permissioned/Permissionless type determines who can join and participate in the network. Anyone can join permissionless networks. In Permissioned networks, nodes need to be allowed by an authority of the network to join in and participate. Bitcoin [4] is an example of a Public/Permissionless network while Quorum [5] and R3 Corda [6] are examples of Private/Permissioned networks.

\subsection{Resource Management Need}
According to The Republic of Turkey Ministry of the Interior Disaster and Emergency Management Presidency Earthquake Office Management's general earthquake statistics report by years in Turkey; Earthquake rates in Turkey are increasing every year. In the last century, 192 damaging earthquakes occurred in Turkey, one hundred thousand citizens died, and more than 650 thousand houses were demolished or damaged heavily [7].

After our researches and interviews, İt's concluded that there is a lack of coordination and resources management in post-crisis management in Turkey. These problems are in the following areas:
\begin{itemize}
\item Coordination between NGOs
\item Management of human resources
\item Resource management
    \begin{itemize}
    \item Logistics
    \item Storage
    \item Distribution
    \end{itemize}
\item Post-disaster necessity tracking
    \begin{itemize}
    \item Verifying the necessity
    \item Support tracking and management
    \item Tracking the provided support 
    \item Preventing abuse of resources and supports
    \end{itemize}
\item Protection of trust in NGOs
\end{itemize}

\subsection{The need for coordination Between NGO’s}

NGOs, which have voluntary activities in disaster situations, have access to different resources such as aid and human resources. It is very important to provide coordination between NGOs to manage these resources correctly. The collaboration of NGOs will enhance the post-disaster recovery.

Autonomous systems can be used to manage these resources effectively and promptly. In this context, utilizing the technology's facilities can have the potential to simplify resource management and ensure coordination.

\section{RELATED WORKS}

There are few studies on disaster management. The use of decision support systems for this purpose is discussed in [8]. Usage of digital twins is given in a recent study [9]. Interestingly, there are comprehensive open-source solutions such as SAHANA EDEN (https://sahanafoundation.org/) that have several functions. Interestingly, Sahane Eden or similar solutions are not that widely used. Our interviews with experts in the field and NGOs revealed some possible reasons for that. It is mainly because these solutions are not built with people in the field. Also, there is always a need for transparency of the transactions and a trust issue in the used systems.

Blockchain technology can be used to overcome these needs. Blockchain technology can remove intermediaries and ensure trust. Blockchain technology was mostly used for cryptocurrency implementations and is famous for high energy usage. However, with the advance in enterprise blockchain frameworks, it is possible to obtain a trusted environment without high energy usage. Different fields of using blockchain are possible such as supply chain, sharing medical information, AI marketplace, DAO machines economy, etc. Also, blockchain can be used for social impact [10].

There are a few decentralized system proposals in the literature regarding this topic. These are mostly for collecting monetary aid [11] or disaster/refugee aid scenarios [12]. An interesting study [13] is about integrating IoT and blockchain in the supply chains of various units to increase the efficiency and effectiveness of humanitarian aid. These papers mostly do not contain technical details and also do not involve NGOs. A very recent study [14] gives a theoretical model which questions the applicability of a trusted supply chain for humanitarian aid. We are not aware of a resource management system for disasters that focuses on NGO collaboration. We focused on generating a prototype to make a social impact.

\section{SYSTEM PROPOSAL}

The proposed NGO-RMSD is shown in Figure 1, nodes to be found in the blockchain network will be added by non-governmental organizations within the system by using blockchain technology. All transactions performed on the system for each node will be saved encrypted in a tamper-proof way. Actors within the distributed application to be created with smart contracts to be deployed on the blockchain network are as follows;
\begin{itemize}
\item Non-Governmental Organizations: Voluntary platforms that are approved by the ministry, supporting local governments and ministries in case of disasters.
\item People In Need: Materially and morally damaged victims due to disasters.
\item Supporter: People, institutions, or organizations that are wishing to provide in-kind and cash support to disaster victims.
\end{itemize}

\begin{figure}[tpb]
\begin{center}
\includegraphics[width=8cm]{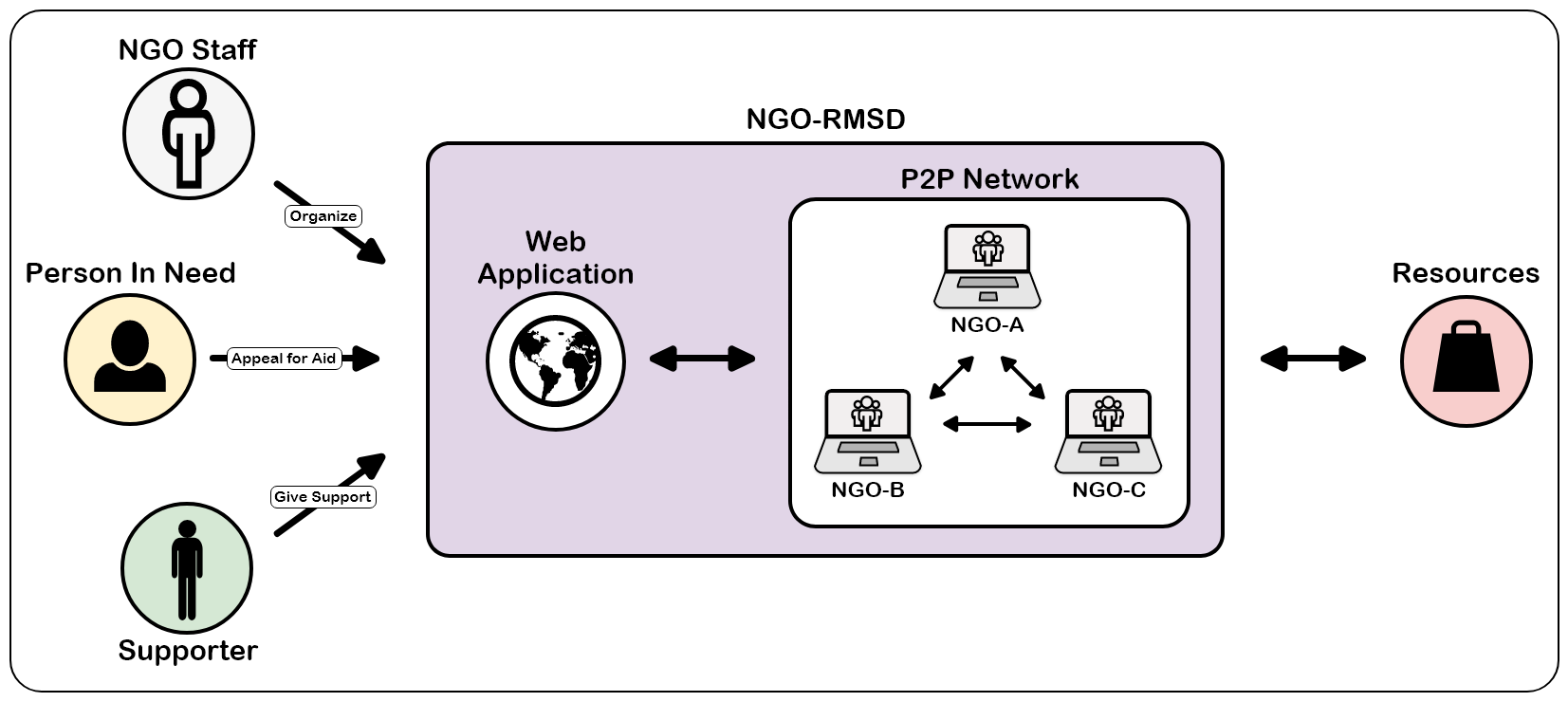}
\end{center}
\caption{Basic design of the NGO-RMSD System}
\label{fig:use_case}
\end{figure}

\subsection{Architecture}
The overall system consists of two parts that are given in Figure 2. A Quorum network for the nodes of the NGO’s and the web service. Those two parts will communicate via Web3JS library from node to JS Back-End. Front-End includes HTML and CSS code to present users from varying platforms with a web page interface. Front-End and Back-End components of the web service will communicate via the ReactJS framework.

To initialize the system, our research group will install the initial (startup) nodes of the network. There is no defined minimum node number, the more the better. The first node account (of the first node) added to the system will have the role “admin”. After the initialization of the system, the account with the admin role will distribute the admin roles to the regarding accounts (pre-defined accounts on the pre-defined nodes) according to the protocol of the community.

At least one node for each NGO is required in the blockchain network to represent the institution. As the system will run in a docker image, there is no need to reserve the node machine only for this purpose. This machine may also be used for general computing purposes by the NGO.

The volunteers or staff of the NGOs are going to create their accounts via those nodes. These accounts are needed to check the support offers or requests and take action. 

RAFT is chosen as the consensus protocol of the proposed system. This protocol uses the “leader-follower” role principle where the leader node is selected internally in the network. The leader and follower nodes can be thought of as minter and verifier nodes in Ethereum. The leader node mines a block to the network while follower nodes verify it. Every node has the chance to be a leader, they have to mine a block for this. Every node has an enode ID that has the raftport parameter as a query string. This parameter will be used to define the communication port that will be used for communicating with the other nodes. These definitions are grouped in the static-nodes.json file. Every node should have the same static-nodes.json file in its filesystem.

The addition of another node or another NGO to the system is as follows:
\begin{enumerate}
\item The candidate NGO node installs the required Docker image.
\item A node can not be initialized as the leader, there is an election for it in RAFT protocol. So every node is added as a new follower node.
\item The required files with the candidate node’s information (static-nodes.json, node keys, etc. ) should be set. The details will be given in the project Github page.
\item The static-nodes.json files of all nodes in the network need to be updated.
\item One node in the network has to run eth.addPeer() command from its geth RPC console with the enode ID of the candidate node. Then, a new node is added and running functionally in the network.
\end{enumerate}

We are currently working on automation for this process and the automatization process will be given on the project Github page.

\begin{figure}[tpb]
\begin{center}
\includegraphics[width=8cm]{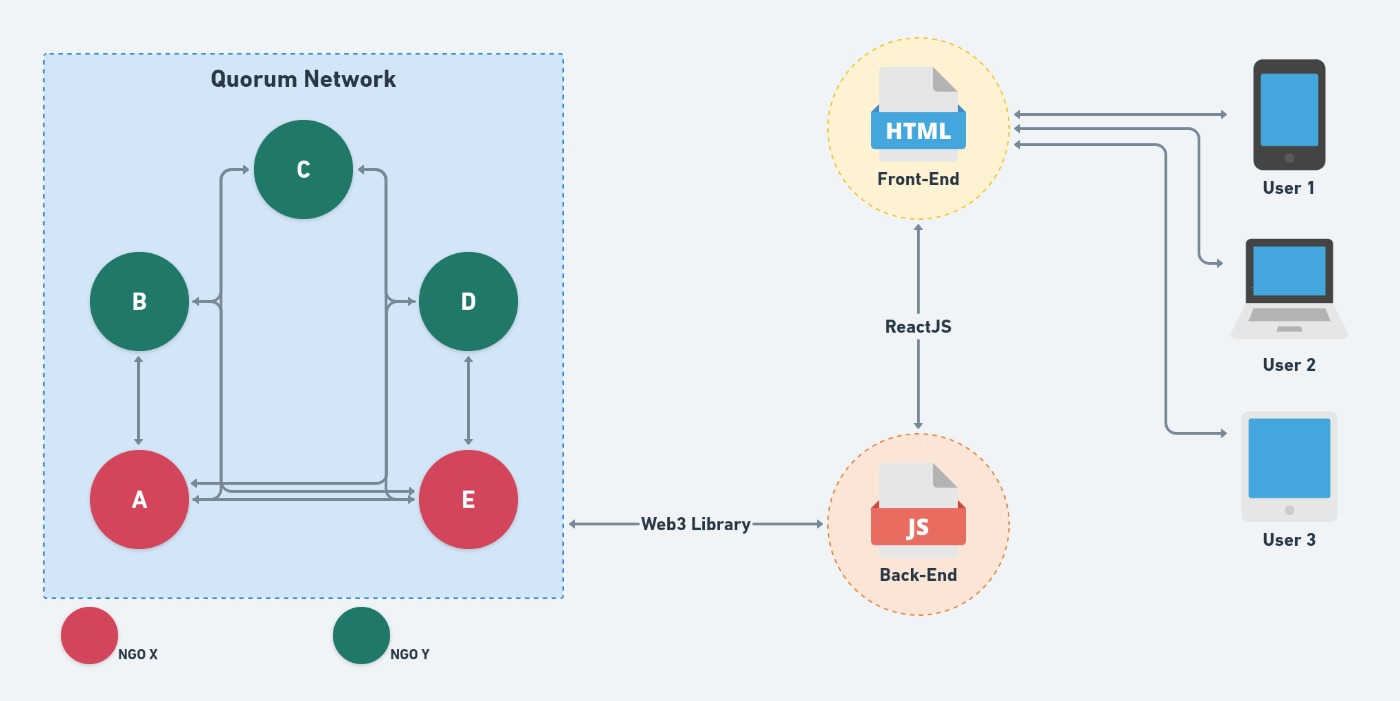}
\end{center}
\caption{NGO-RMSD System Architecture}
\label{fig:system_architecture}
\end{figure}

\subsection{Smart Contracts}

Although the blockchain concept is said to be decentralized, not all solutions have to be like that. Private/Permissioned networks like Quorum are not fully decentralized. The system gives privileges to some authorities to add nodes or users. 

The proposed system aims to form trust and transparency. NGOs have the ADMIN role which has the privileges that can set the user's roles as CHECKER or CREATOR. Also, NGOs with the ADMIN privileges add the new NGOs to the system. Users are assigned roles in the smart contract structure. These roles are as follows: 
\begin{enumerate}
\item Admin: Admin is the system administrator. 
\item Checker: The role of the checker belongs to NGOs and is used to check whether the proposed support or request is correct.
\item Creator: The role of the creator is open to everyone and it is required to make a request or support offered.
\end{enumerate}

Smart contracts are written for the following functions:
\begin{enumerate}
\item Role Assignment and Listing Functions: These are called by the system administrator and they aim to assign the role to the user with the setUser function and to display the role of the user with the getUserAuth function.
\item Authority-Role Control Modifier Functions: These functions are called by the organizer to compare the hash of the user's role with the hash of the role assigned to the user.
\item Requirement Creation Function: This function enables users in the role of “Creator” to create a new need with the data that they enter. Need; is created with features such as type, and amount. The created need is put in the need list and the label "waiting for confirmation" is added.
\item Support Creation Function: This function enables users in the role of "Creator" to create new support with the data that they enter. Support; is created with features such as type, amount, and shipping type. The created support is added to the support list and the label "waiting for approval" is added.
\item Approve Functions: These functions enable users with the role of "Checker" to confirm the needs and supports labeled "waiting for approval". While the approveNeed function confirms the pending needs, the approveSupport function confirms the pending supports. Once the need is approved, the label changes to "approved" and not "waiting for approval". Once the support has been approved, it is added to the approved support list. 
\item  Request-Support Listing Functions: These functions are used to display all needs and supports. While the showSupport function listed the known support, the showSupports function shows all the supports. The showAllApprovedSupports function lists approved supports. The showNeedOffers function lists all the needs records while the showNeed function lists the known need records. 
\end{enumerate}

The last 2 functions (showNeedStatus and showSupportStatus) could be run by users with the role of "Checker", and these are in charge to list the known requests and supports confirmation status.

\subsection{System Application Flow}
Application flow of the system as shown in Figure 3, the user chooses their application type. That application is sent to the Quorum node and stored as an unapproved application. A transaction is sent to the network and the ledgers of nodes get updated. If there are any unapproved applications in the network, an NGO Staff checks the application by communicating with the user that applied. If the application is valid, in that case, the requirement is real or support is accurate; the information of the applicant user is written into the local database, and the application gets approved. After the application validation, the application is stored as an approved application. Then, another transaction is sent to the network, and the ledgers of nodes get updated again. If there are no unapproved applications, the application flow process is finished.

\begin{figure}[tpb]
\begin{center}
\includegraphics[width=8cm]{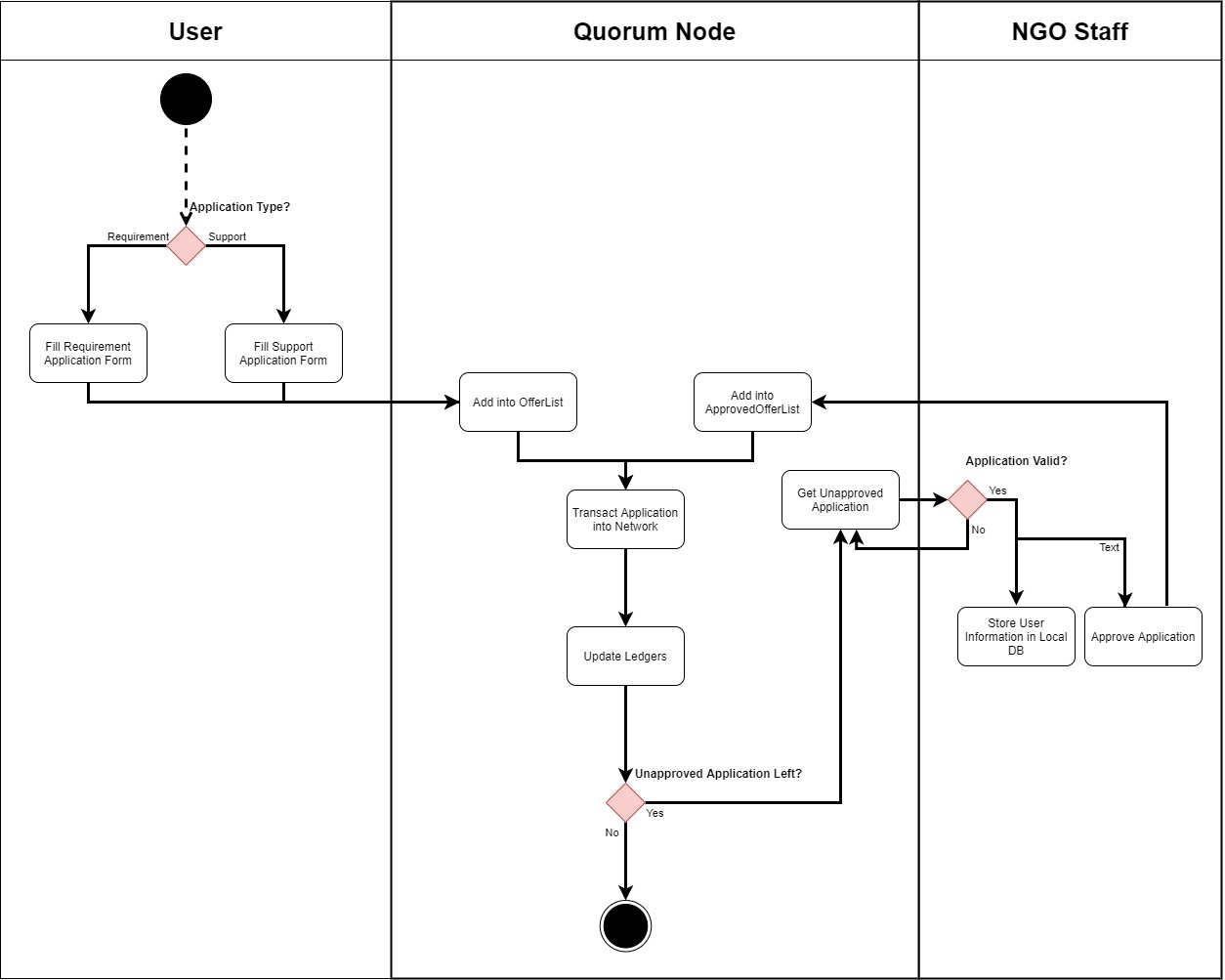}
\end{center}
\caption{System Activity Diagram}
\label{fig:activity_diagram}
\end{figure}

\section{IMPLEMENTATION}
The default system works initially as three nodes with two NGOs. To participate in the network and the system, each NGO must add at least one node. The scenario with five nodes was executed. The proof of concept implementation of NGO-RMSD is implemented on the Quorum framework. The Quorum framework, which provides corporate solutions, is used as blockchain technology. It offers a private/permissioned blockchain structure with low energy consumption [5]. Docker container technology is used to integrate this system into every platform. The system does not need special hardware and does not need high power. Ordinary virtual machines (single CPU, 4 GB ram, 256 GB disk) are used for the nodes. 

Web3JS is used for the communication between the Quorum network and Web interface backend. Solidity language is used to write smart contracts. RAFT protocol is used as the consensus protocol as it can support more nodes than Istanbul Byzantine Fault Tolerance (IBFT). NodeJS is used in the backend because it provides the asynchronous functionality that this application needs.

The privacy of personal data is prioritized in the design. Turkish Law on the Protection of Personal Data (KVKK) [15] and the EU General Data Protection Rules (GDPR) [16] requirements are taken into consideration. Since it is not possible to delete the data on the blockchain, the storage of personal data in the blockchain constitutes a situation that is not suitable for personal data protection laws. In our design, personal data will never be stored on the blockchain. Personal data will be stored in a private database instead, which could be deleted at any time by NGOs. In a privacy breach case, it will be the NGO's responsibility if it didn't take the required actions to apply to the necessary legal processes. However, Non-interactive zero-knowledge proof (NIZK) based autonomous codes can be developed to ensure privacy [17], which is out of the scope of this study.

In this project, open-source and free software licensed software is used, and dependency on any company or institution is minimized. This project is also published on the Github page.

The prototype is created appropriately with the scenario specified. Requirement Creation, Requirement Confirmation, Support Creation, Support Confirmation, Support, and Requirement Listing processes are directed to smart contracts running on nodes in the Quorum network through the web interface.

\subsection{Smart Contracts}

The smart contract is open to improvement and updating and debugging process continues. The recent version of the contract structure and functions are given on the public Github page \url{https://github.com/MSKU-BcRG/akys}.

\subsection{GUI Design}
Prototype web interfaces of the system are designed. Disaster victims can make requests or give support through forms that are shown in Figure 4. All operations in the system are stored on the blockchain as transactions. All the users can track the status of the transactions through the web interface that is given in Figure 5.

\begin{figure}[tpb]
\begin{center}
\includegraphics[width=8cm]{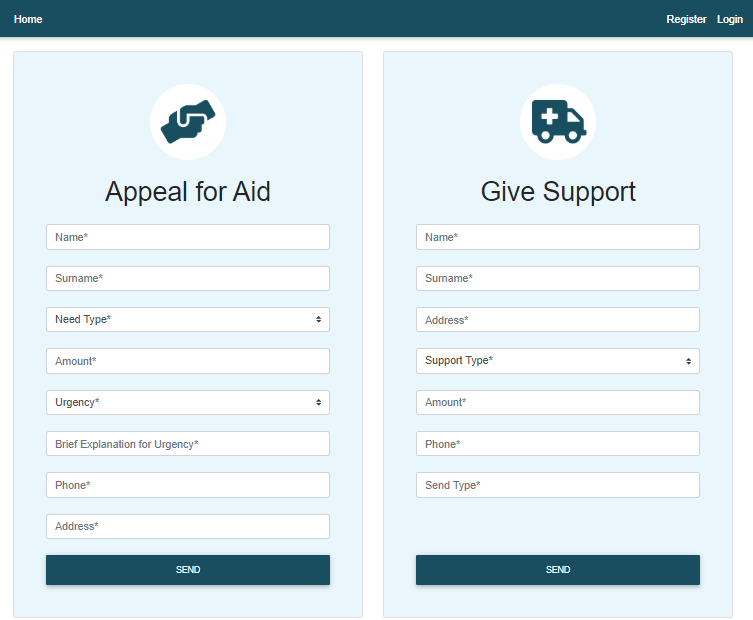}
\end{center}
\caption{Request and Support Forms}
\label{fig:gui1}
\end{figure}

\begin{figure}[tpb]
\begin{center}
\includegraphics[width=8cm]{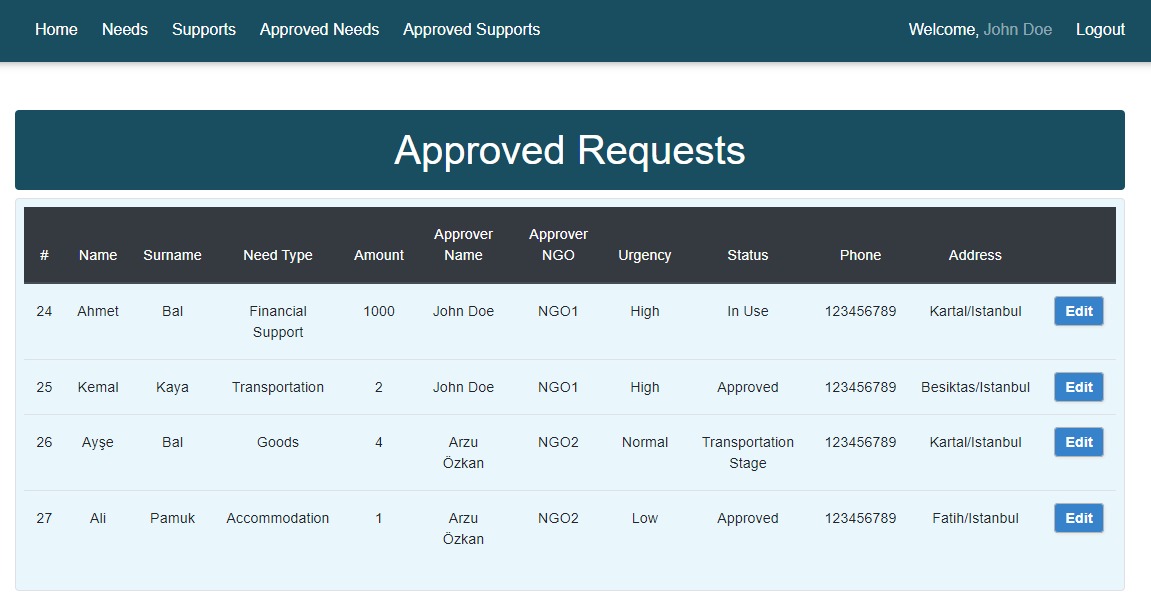}
\end{center}
\caption{Interface to track approved requests}
\label{fig:gui2}
\end{figure}

\section{ACKNOWLEDGEMENT}
This study won the second prize in the TUBİTAK 2242 University Students Research Project Competition in the field of “Information and Communication Technologies” in 2020.

\section{RESULTS AND CONCLUSION}
Decentralized resource management and NGO coordination system are proposed for disasters in this study. The proposed NGO-RMSD system will enable NGOs and public institutions to act in coordination in the case of disasters. A trusted system is designed where all transactions are transparent. Proof of concept implementation of the proposed system NGO-RMSD had promising results. Smart contracts which allow the autonomous working of the system is written and tested on the live platform. 

Coordination of NGOs will be provided with this system which will lead to improvements in workforce and resource management. The proposed system will enable NGOs to reach more people in need. Also, it will ensure urgent needs could be determined to resolve as soon as possible. Users' personal data's unauthorized sharing will be prevented and made sure that the data-sharing is legal. 

The smart contracts and project details of the system are shared on the project's Github page with a free software license. We are in contact with related NGOs and further development with the community will continue. The live tests of the proposed system is being implemented in the DS4H blockchain research network. Dynamically establishing new nodes and the automation for the authentication will be given on the project's Github page. Details of the project can be reached at \url{http://wiki.netseclab.mu.edu.tr/index.php?title=STK-AKYS}.

In future works, it is aimed to ensure that the system is easily integrated into new laws and regulations. The personal data will be protected autonomously with new smart contracts. It is also aimed to create a more efficient system by integrating the data collected in the system with the data collected from the territory. The collected data will be processed with artificial intelligence.  The requirement and support requests will be matched by using natural language processing and machine learning. A dataset for machine learning processes will be formed with the collected data on the live system. Thus, the data to be entered into the system could be cleared by using Natural Language Processing (NLP)  and a classification algorithm could be applied with keywords. It will be possible to apply the need and support matching infrastructure through this classification algorithm. In all these processes, preliminary studies will be done to receive extra information from the Twitter environment with sustained queries to be made with these keywords.

\vspace{12pt}
\color{red}

\end{document}